\documentclass[journal = jpccck]{achemso}
\usepackage{natbib}
\usepackage{graphics,graphicx}
\usepackage[dvipsnames]{xcolor}
\usepackage{tabularx}
\usepackage{chemformula}
\usepackage[T1]{fontenc}
\usepackage{cancel}
\usepackage{times}
\usepackage{amssymb,amsfonts,amsmath}
\usepackage{soul}
\usepackage{threeparttable}
\usepackage{multirow}

\newlength{\figurewidth}
\usepackage[version=4]{mhchem}

\newlength{\figwidthcol}
\newlength{\figwidthtwocol}
\def\eg{\textit{e.g.}}
\def\ie{\textit{i.e.}}

\usepackage{color}

\title{Structurally Constrained Evolutionary Algorithm for the Discovery and  Design of Metastable Phases}

\author{Busheng Wang}
\affiliation{Department of Chemistry, State University of New York at Buffalo, Buffalo, NY 14260-3000, USA}
\author{Katerina P. Hilleke}
\affiliation{Department of Chemistry, State University of New York at Buffalo, Buffalo, NY 14260-3000, USA}
\author{Samad Hajinazar}
\affiliation{Department of Chemistry, State University of New York at Buffalo, Buffalo, NY 14260-3000, USA}
\author{Gilles Frapper}
\affiliation{Applied Quantum Chemistry Group, E4 Team, IC2MP UMR 7285, Université de Poitiers-CNRS, 86073, Poitiers, France}
\author{Eva Zurek}
\email{ezurek@buffalo.edu}
\affiliation{Department of Chemistry, State University of New York at Buffalo, Buffalo, NY 14260-3000, USA}

\begin{document}
\begin{abstract}
Metastable materials are abundant in nature and technology, showcasing remarkable properties that inspire innovative materials design. However, traditional crystal structure prediction methods, which rely solely on energetic factors to determine a structure's fitness, are not suitable for predicting the vast number of potentially synthesizable phases that represent a local minimum corresponding to a state in thermodynamic equilibrium. Here, we present a new approach for the prediction of metastable phases with specific structural features, and interface this method with the \textsc{XtalOpt} evolutionary algorithm. Our method relies on structural features that include the local crystalline order (\eg, the coordination number or chemical environment), and symmetry (\eg, Bravais lattice and space group) to filter the breeding pool of an evolutionary crystal structure search. The effectiveness of this approach is benchmarked on three known metastable systems: XeN$_8$, with a two-dimensional polymeric nitrogen sublattice, brookite TiO$_2$, and a high pressure BaH$_4$ phase that was recently characterized. Additionally, a newly predicted metastable melaminate salt, $P$-1 WC$_{3}$N$_{6}$, was found to possess an energy that is lower than two phases proposed in a recent computational study. The method presented here could help in identifying the structures of compounds that have already been synthesized, and developing new synthesis targets with desired properties. 
\end{abstract}

\maketitle
\newpage
\section{Introduction}
Thermodynamically metastable phases, which are kinetically trapped and do not correspond to the global minimum in the free energy landscape, are commonly found in nature and can be synthesized in the laboratory and mass produced industrially~\cite{sun2016thermodynamic}. According to the Materials Project database (accessed in June 2023), 53.13\% of the experimentally verified phases are metastable. \cite{jain2013commentary} These compounds can exhibit a range of unique properties, which can be attributed to a larger number of possible atomic arrangements and chemical bonding configurations they can adopt as compared to their ground states. Just a few examples of the broad spectrum of applications that metastable phases may be used for include superconductors~\cite{budden2021evidence,yoshida2018metastable}, photocatalysts~\cite{sclafani1996comparison}, photovoltaics~\cite{nagabhushana2016direct, kuech2016growth}, ion conductors~\cite{sanna2015enhancement}, and steels~\cite{li2016metastable}. These examples illustrate that metastable phases may present novel opportunities for materials innovation and design.

In the last decade computational crystal structure prediction (CSP) has become a powerful tool facilitating materials discovery~\cite{oganov2011modern,zurek2015predicting,oganov2019structure}. A number of methods have been developed and applied to predict the structures of materials without input from any experimental data including random search~\cite{pickard2009structures}, simulated annealing~\cite{kirkpatrick1983optimization}, metadynamics~\cite{martovnak2003predicting}, minima hopping~\cite{goedecker2004minima}, basin hopping~\cite{wales1997global,burnham2019crystal,yang2021exploration}, particle swarm optimization~\cite{wang2010crystal}, evolutionary algorithms~\cite{oganov2006crystal,trimarchi2007global,lonie2011xtalopt,tipton2013grand,hajinazar2021maise,wu2013adaptive}, as well as template-based elemental substitution~\cite{wei2022tcsp}. Traditionally, these methods have been devoted towards finding the global (free) energy minimum (ground state), and predicting the crystal structure for a desired metastable material -- at best, a secondary target -- has remained a formidable challenge. The reason for this is that uncovering the structures of low-lying metastable phases that could potentially be synthesized is complicated by the vast number of local minima that are not typically accessed using traditional CSP methods.

In recent years, significant progress has been made in advancing metastable phase prediction through data-assisted CSP searches. Various computed physical and chemical properties have been employed as objectives to be optimized (either in lieu of, or in addition to the enthalpy/energy) including but not limited to hardness \cite{lyakhov2011evolutionary,zhang2013first,avery2019predicting}, density \cite{zhu2011denser}, band gap \cite{xiang2013towards,zeng2014evolutionary}, the degree of interstitial electron localization\cite{zhang2017computer}, phase transitions~\cite{zhu2015generalized},  charge-carrier mobility \cite{cheng2020evolutionary}, singlet fission rate \cite{tom2023inverse}, as well as structural information (\eg, predefined molecular structures \cite{zhu2012constrained,curtis2018gator,case2016convergence,budden2021evidence,kilgour2023geometric}, high-symmetry space groups \cite{wales1997global,wheeler2007sass,huber2023targeting}).  Experimental data (\eg, comparison of simulated and experimental X-ray diffraction (XRD) patterns~\cite{gao2017x,ward2017automated,ling2022solving}) have also been employed within such CSP searches. 

Yet, comparison of computed observables with experiment may, at times, be problematic. This is especially true when the structural characteristics of a synthesized compound cannot be fully measured and only partial geometric details can be determined. Examples of data that are helpful, but not sufficient, to characterize a particular crystal include local order or coordination environments that can be probed using infrared and Raman spectroscopy~\cite{gard1986infrared,wang2020s2}, or the symmetry of a particular lattice or sub-lattice obtained via XRD~\cite{pena2022chemically,dasenbrock2023evidence}. In such cases, the absence of complete structural information can impede precise compound characterization. For example, determining the crystal structure of hydrides is a particularly challenging task,\cite{einaga2016crystal,pena2022chemically,dasenbrock2023evidence} as hydrogen provides weak X-ray scattering in contrast to heavier metallic atoms, given the proportionality of the scattering power to the atomic number~\cite{schmidtmann2014determining,cheng2017synchrotron}. Additionally, when it is known that specific structural features are beneficial for a particular property, these can be leveraged in the design of hypothetical materials. Some examples include thermodynamically metastable oxalate-based Li$_{x}$(CO$_{2}$)$_{y}$ compounds that can be employed as organic materials for lithium-ion batteries~\cite{huang2018pressure}. The impact of such theoretical insights cannot be understated, as they offer valuable avenues for expanding the design possibilities of important materials in a forward-thinking manner. Therefore, the development of a versatile global search method for identifying crystal structures of metastable phases that incorporates partial structural features derived from theoretical models and experimental data is highly desirable. 

In this study, we present a new approach for determining the crystal structure of a metastable phase with specific structural features based on the \textsc{XtalOpt}~\cite{lonie2011xtalopt,falls2020xtalopt} evolutionary algorithm (EA). Our method automatically filters out structures that do not adhere to predefined criteria including but not limited to the local crystalline order (\eg, coordination number and chemical environment) and symmetry (\eg, Bravais lattice and space group)  before ranking according to total energy or enthalpy. The accuracy and effectiveness of this method was evaluated by benchmarking its performance against three systems: XeN$_{8}$\cite{wang2022putting} containing 2D polymeric nitrogen motifs at 50~GPa, brookite TiO$_{2}$ at ambient pressure, 
and the BaH$_{4}$ phase synthesized at 50~GPa that possesses an $I$4$/mmm$ Ba sub-lattice \cite{pena2022chemically}. Additionally, we have predicted a new metastable melaminate salt, $P$-1 WC$_{3}$N$_{6}$, which is found to be lower in energy compared to the compounds reported recently by Chen, Wang and Dronskowski~\cite{chen2023computational}. Our results demonstrate that a constrained search is more efficient in identifying metastable phases with user-defined structural features compared to a traditional CSP search. 

\section{Methods}
\textit{Structure prediction}. Our CSP searches were carried out using the in-house developed \textsc{XtalOpt}~\cite{lonie2011xtalopt, falls2020xtalopt} EA (release 12)~\cite{avery2019xtalopt}, which was designed to find stable and metastable structures of crystals given only their composition. \textsc{Pymatgen}~\cite{ong2013python}, an open-source Python library for materials analysis, was utilized to determine the degree of local crystalline order (\eg, coordination number, bond lengths and chemical environments) and symmetry (\eg, space group and Bravais lattice) as global variables for each structure that could be used in a constrained search for metastable phases, by restricting the breeding pool\cite{zurek2016discovering} to crystals that possessed the desired structural feature. Detailed information regarding the specific functions/classes of \textsc{Pymatgen} and scripts utilized for applying the constraints are provided in the Supporting Information (SI). In this study, the symmetry analyzer method within \textsc{Pymatgen} was implemented with a symmetry tolerance of 0.02 for both BaH$_{4}$ and TiO$_{2}$. Initial tests showed that a symmetry tolerance value of 0.02 is appropriate for both cases. A cutoff value of 1.5 \AA{} was used for the N-N and C-N bond distance to identify the neighbor atoms, see Section S1 in the SI, for more detailed information on how this value was chosen. The energies or enthalpies associated with the phases that did not adhere to the structural constraints were assigned an exceptionally large value (\eg, 99~eV per formula unit or FU). Because the breeding pool was chosen to include only the 50 most stable structures, this artificially large value prevented the structures that did not satisfy the constraints from populating the breeding pool. The CSP searches for XeN$_{8}$ (1-2), TiO$_{2}$ (2, 4, 8), BaH$_{4}$ (2, 4), and WC$_{3}$N$_{6}$ (1-2) were conducted with the number of FU in the primitive cell given in the brackets.  PyXtal \cite{fredericks2021pyxtal} was used to generate molecular crystals as seeds for WC$_{3}$N$_{6}$, which contains melaminate, (C$_{3}$N$_{6}$)$^{6-}$, motifs and tungsten atoms.

\textit{First-principles calculations}. The geometry optimizations and electronic structure calculations were performed using the Vienna \textit{Ab Initio} Simulation Package (\textsc{VASP})~\cite{kresse1996efficient,kresse1996efficiency}, coupled with the Perdew-Burke-Ernzerhof (PBE) gradient-corrected exchange and correlation functional~\cite{perdew1996generalized}, which was used for all of the CSP searches. To test the effect of dispersion interactions on the relative energies or enthalpies, the  D3 (BJ)\cite{grimme2010consistent,grimme2011effect} and optB88\cite{klimevs2009chemical} approaches were used in the case of WC$_{3}$N$_{6}$ and XeN$_8$. The electron-ion interaction was described by the projector-augmented-wave method~\cite{blochl1994projector}, in which the Xe~4$d^{10}$5$s^2$5$p^6$, N 2$s^2$2$p^3$, Ti 3$d^3$4$s^1$, O 2$s^2$2$p^4$, Ba 5$s^2$5$p^8$6$s^2$, H 1$s^1$, W 5$d^5$6$s^1$, and C 2$s^2$2$p^2$ configurations were treated as valence. In the structure searches, the geometries were optimized with the ``accurate'' precision setting in VASP using a three-step process. First, the atomic positions were allowed to relax within a set unit cell, which was followed by a volume-only relaxation. Lastly, a full relaxation was conducted. A plane-wave basis set with an energy cutoff of 520 eV, and a $\Gamma$-centered Monkhorst-Pack $k$-mesh~\cite{monkhorst1976special} where the number of divisions along each reciprocal-lattice vector was chosen such that its product with the real lattice constant was 30~\AA{}, was employed. 
In order to conduct precise reoptimizations and electronic structure calculations, we increased the cutoff values to 600 eV for TiO$_{2}$ and WC$_{3}$N$_{6}$ under ambient pressure. For XeN$_{8}$ and BaH$_{4}$ under pressure, the cutoff values were set at 1000 and 700 eV, respectively. Additionally, we augmented the $k$-mesh value to ensure that its product with the real lattice constant was 50~\AA{}. The phonon spectra of the new phase of WC$_{3}$N$_{6}$ was computed using the frozen phonon method implemented in the \textsc{PHONOPY}\cite{togo2015first} code.

\section{Results and Discussion}
\subsection{Constrained Evolutionary Search -- Workflow}
Our methodology for setting up the constrained CSP searches, which are described in this manuscript, is built upon the open-source EA, \textsc{XtalOpt}~\cite{falls2020xtalopt}, developed in our group. In its most basic form, \textsc{XtalOpt} constructs possible unit cells and atomic coordinates of structures that sample the full 3$N + $3 (3$N - $3 for the atomic positions, 3 for the unit cell angles and 3 for the lattice vectors) dimensional potential energy surface (PES), and are subsequently relaxed to the nearest stationary point via an external program. Most often, a periodic density functional theory (DFT) package is employed, though codes that make use of interatomic potentials have also been interfaced with \textsc{XtalOpt}. The goal of traditional \textsc{XtalOpt} searches is to find the location in the PES that corresponds to a minimum in the energy or enthalpy, termed the global minimum. Without including effects such as temperature, configurational entropy or reaction barriers, this would correspond to the structure expected to be dominant in a system at chemical equilibrium. In addition to the global minimum, many local minima separated by energy barriers are scattered throughout the PES, which correspond to alternative potential configurations of atoms in unit cells, albeit ones that are higher in energy than the global minimum. Along the way, \textsc{XtalOpt} may also identify structures at these local minima, although only the global minimum is explicitly targeted. The local minima represent metastable structures, which does not preclude them from experimental observation nor from persisting once formed. Diamond is metastable at ambient conditions, representing a local minimum in the PES of elemental carbon, whereas graphite is the global minimum. Regardless of its metastable status, the diamond allotrope of carbon still -- famously -- persists ``forever'' at ambient temperature/pressure conditions.

In the same way that evolution in a biological population adapts to its environment over time, with favorable traits being passed down from generation to generation, the \textsc{XtalOpt} EA will selectively choose favorable (low-energy) configurations for propagation to future generations of structures. The ``fitness" of each of these configurations is evaluated, typically according to their thermodynamic stability represented by enthalpy or energy. This procedure must be modified to target local rather than global minima. Therefore, in the current work we introduce additional criteria, by constraining the crystals that make up the breeding pool to those that contain user defined structural features. In such a way the selection, mutation, and recombination operators will work on the population of atomic configurations \emph{that match the structural criteria} to generate the lowest-energy structures that possess the desired motifs.  This approach effectively enhances the accuracy and efficiency in predicting metastable phases that contain user-defined structural features.

\begin{figure}[]
\centerline{\includegraphics[width=0.5\columnwidth]{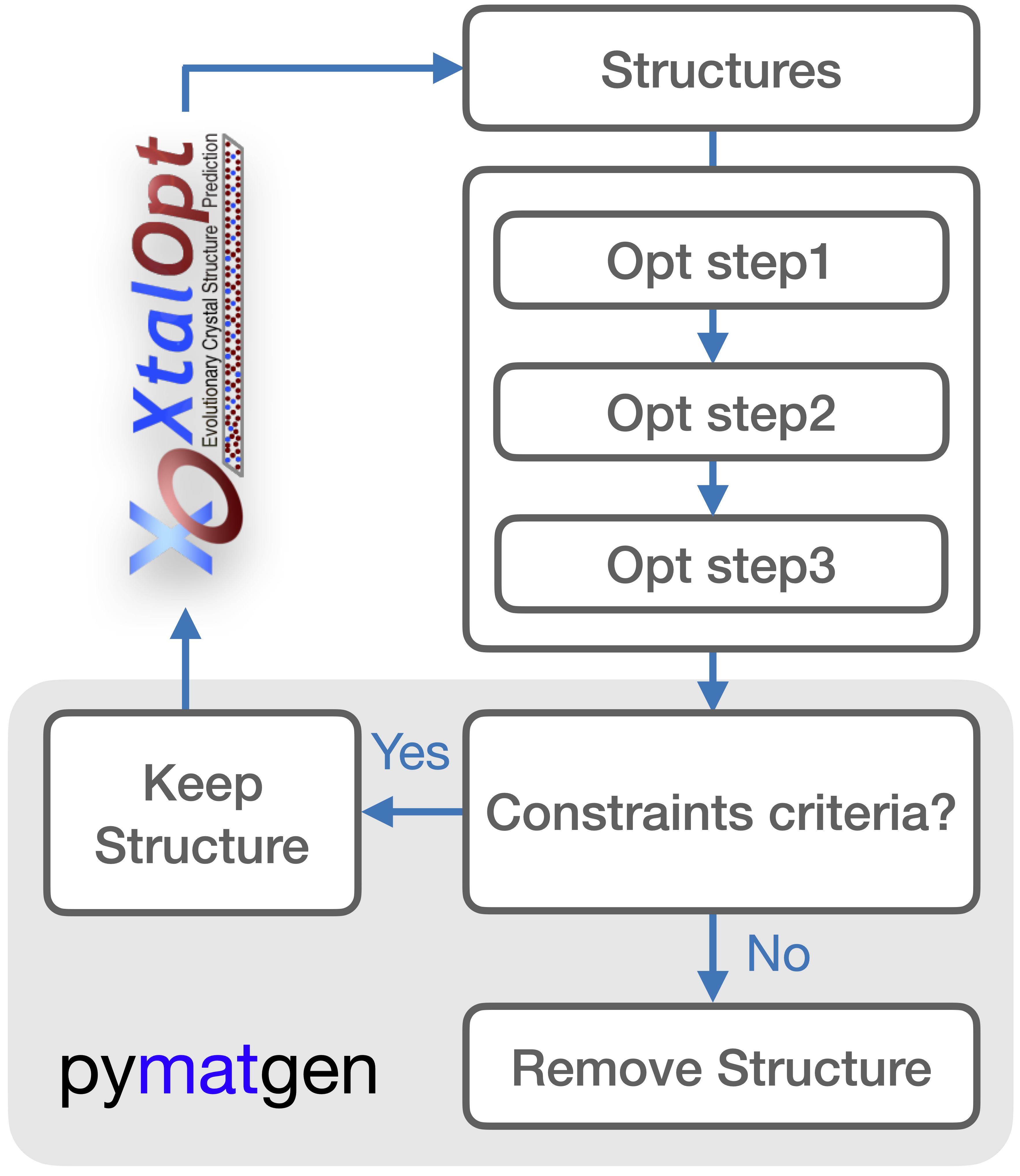}}
\caption{Schematic workflow for conducting constrained crystal structure prediction searches through the utilization of the \textsc{XtalOpt} EA coupled with the \textsc{Pymatgen} library. Structures are generated as in a traditional \textsc{XtalOpt} search (open area), then passed to an external program for local relaxation (white box). The relaxed structures are analyzed via \textsc{Pymatgen} (grey box), and the breeding pool upon which \textsc{XtalOpt} can apply evolutionary operators is filtered. The constraints, which may include specified coordination numbers and bond lengths, space group symmetry or Bravais lattice, are applied after the geometry optimizations are complete, and these constraints remove unsuitable structures from the breeding pool\cite{zurek2016discovering} (see Figure S1 in the SI for further details on the workflow of a traditional \textsc{XtalOpt} search, which includes determination of a breeding pool).}
\label{fig:workflow}
\end{figure}

Figure \ref{fig:workflow} shows a schematic workflow for conducting constrained searches through the utilization of \textsc{XtalOpt} coupled with the \textsc{Pymatgen} library. As in a traditional \textsc{XtalOpt} search the initial set of structures may either be seeded by the user, or generated randomly, for example via the \textsc{RandSpg} algorithm \cite{Zurek:2016h} that can be employed to create random symmetric crystals. Next, the geometries of these initial structures may be optimized loosely using any of the total energy calculation methods available in the \textsc{XtalOpt} code. In the present work, the local relaxations were performed by first-principles DFT in a three-stage optimization step as described in the \emph{Methods} section above. This reduces the noise of the potential energy surfaces and favors the generation of chemically sensible structures to accelerate the speed at which favorable geometries are identified. Post-optimization, duplicates are removed from the breeding pool using the \textsc{XtalComp} algorithm \cite{Zurek:2011i} to maintain structural diversity. What is new in this work is that the crystals in the breeding pool are further subject to structural constraints that are applied via the \textsc{Pymatgen} library\cite{ong2013python}, as described above. Crystals that do not meet the specified constraints are removed from the breeding pool. This exclusion occurs only after structures have been optimized with DFT because the initial geometries that are made using evolutionary operators typically have no symmetry and have interatomic distances that may noticeably differ from those in the optimized systems. Next, as in a traditional search, the evolutionary operators are applied to the optimized structures (parents) to generate the child structures.

In what follows, we provide examples of the utility of this method on a number of systems chosen to illustrate how appropriately defining the desired coordination numbers, chemical environments, as well as the space group and Bravais lattice type can be used to filter the breeding pool. The constrained technique  described in Figure \ref{fig:workflow} has been integrated with the \textsc{XtalOpt} software, and we demonstrate its utility for XeN$_{8}$, and brookite TiO$_{2}$, which are metastable by 102 and 13~meV/atom, respectively, at 50 GPa and ambient pressure. Furthermore, we also used the method to predict the structure of a synthesized high-pressure phase of BaH$_{4}$, whose Ba sublattice was characterized with the aid of XRD \cite{pena2022chemically}. Finally, utilizing the methodology proposed herein a new metastable H-free melanine phase, W(C$_{3}$N$_{6}$), is predicted. This polymorph is calculated to be lower in energy than the previously reported potential multifunctional materials based upon the (C$_{3}$N$_{6}$)$^{6-}$ unit~\cite{chen2023computational}.

\subsection{Constraining Coordination Numbers for Prediction of 2D Polymeric Nitrogen}
The prediction and discovery of novel polymeric nitrogen or nitrogen-rich compounds is of great interest in contemporary times because of their potential as environmentally-friendly high-energy density materials (HEDMs). Such compounds should ideally possess N-N single or N=N double bonds, which may release a vast amount of energy upon decomposition to the inert and thermodynamically stable \ch{N+N} triple bond found in N$_2$. Several polymeric nitrogen configurations have been previously reported such as chain\cite{mattson2004prediction}, layered\cite{tomasino2014pressure,laniel2019hexagonal,laniel2020high,wang2010structural}, cubic-gauche allotrope (cg-N)\cite{eremets2004single}, cage-like\cite{sun2013stable}, N$_{5}$, N$_{6}$, and N$_{8}$ molecular crystals\cite{huang2021predicted,wang2020prediction,greschner2016new,hirshberg2014calculations}. Note that four of these computationally predicted phases have been experimentally characterized, \ie, bulk cg-N\cite{eremets2004single}, layered polymeric nitrogen (LP-N)\cite{tomasino2014pressure}, a layered black-phosphorus-type nitrogen allotrope (BP-N)\cite{laniel2019hexagonal}, and hexagonal layered polymeric nitrogen (HLP-N)\cite{laniel2020high}. Contrary to LP-N, BP-N, and HLP-N, which are composed of 3-coordinated nitrogen atoms and possess layers that are at least two atoms thick, a novel polymeric nitrogen with layers a single layer thick has recently been predicted, with intercalated alkali metal (K\cite{steele2017novel}) or noble gas (Xe\cite{wang2022putting}) atoms. This new family of structures characterized by two-dimensional (2D) polymeric nitrogen possesses crown ether-like nanopores, which contain both 2- and 3-coordinated N atoms.

\begin{figure}[]
\centerline{\includegraphics[width=1.0\columnwidth]{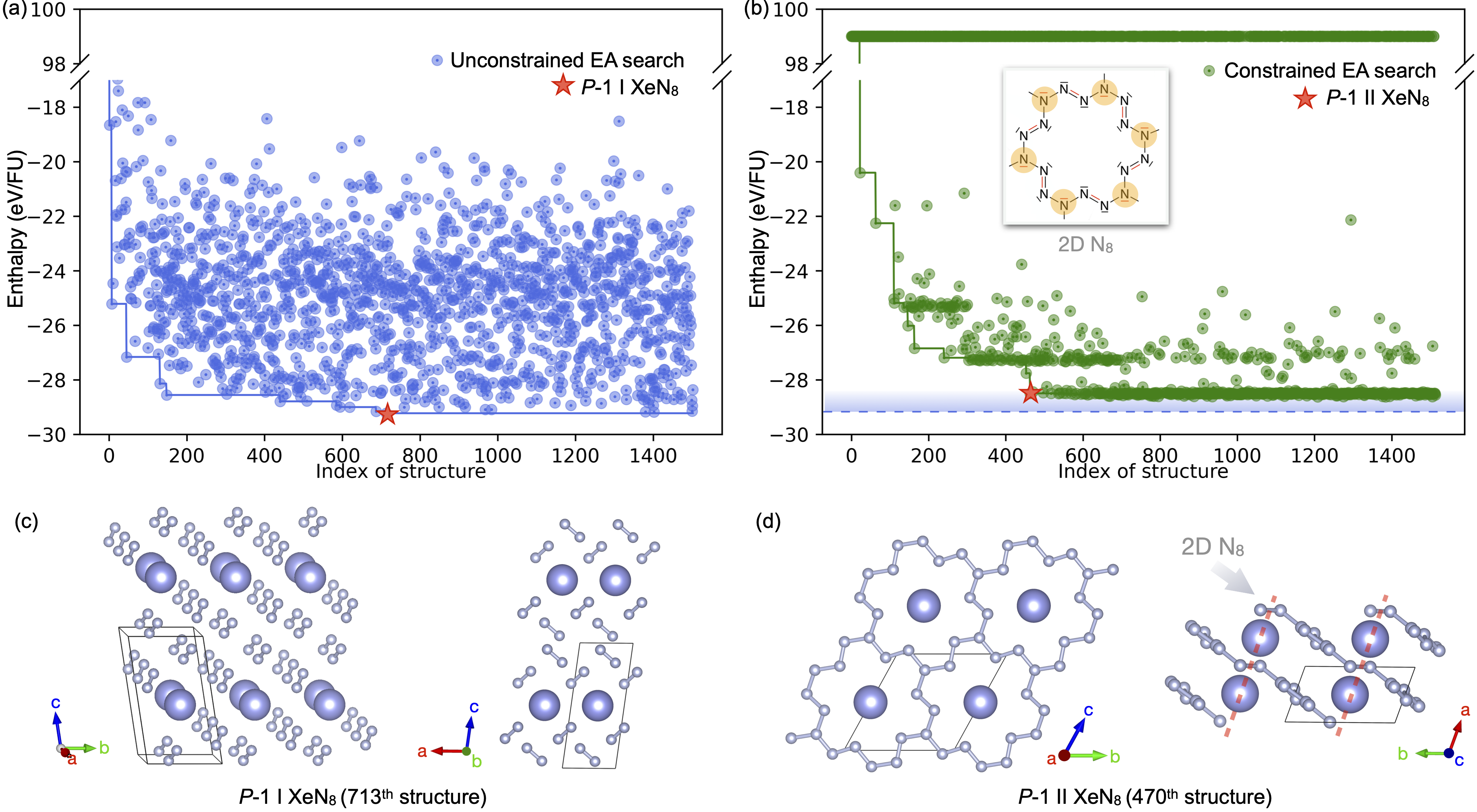}}
\caption{(a-b) The evolution of the enthalpy versus the structure index for the XeN$_8$ stoichiometry at 50~GPa performed with a CSP search without (a) and with (b) constraints is shown by the blue and green dots, respectively. Approximately 1500 crystals were optimized in both runs.  The most stable structures found within these searches are indicated by the red stars. (c-d) The crystal structures of $P$-1 I (the global minimum) and metastable $P$-1 II XeN$_{8}$. The Lewis structure of the 18-crown-6 N$_{18}$ unit in the 2D-N$_{8}$ layer of $P$-1 II XeN$_{8}$ is shown in (b). The solid thin blue and green lines in (a) and (b) denote the lowest enthalpy structure found during the course of the search. In (b) the blue dashed line represents the enthalpy of the ground state $P$-1 I XeN$_{8}$ phase. The enthalpies of the discarded phases in the constrained search are represented with a constant value of 99~eV/FU. Nitrogen atoms are small and xenon atoms are big balls.}
\label{fig:xen8}
\end{figure}

To illustrate the evolution of a CSP search carried out with the traditional \textsc{XtalOpt} algorithm versus one performed using constraints, it is useful to plot the DFT energies or enthalpies of the optimized crystals versus their structure index, which is directly related to the time at which the structure was generated during the course of the run. Figures \ref{fig:xen8}(a) and (b) illustrate the progress of the unconstrained and constrained searches, respectively, carried out on the XeN$_{8}$ stoichiometry  at 50~GPa. Before describing the exact constraints that were employed, let us comment on how these two plots differ. In the constrained search, the discarded structures (denoted using a thick line at 99~eV/FU) corresponded to 57.4\% of all optimized phases, and were deemed ineffective structures that have been excluded from consideration. From the remaining crystals, 99.1\% of the low-lying structures were located within the range of -25.0 to -29.0 eV/FU, as shown in Figure \ref{fig:xen8}(b).  Meanwhile, it is evident that the majority of these effective structures can be broadly categorized into three distinct enthalpy regions: -25.0 to -25.5~eV/FU, -27.0 to -27.5~eV/FU, and -28.3 to -29.0~eV/FU. The structures falling within these specific enthalpy ranges contain the desired 2D polymeric nitrogen, with the primary distinction between them being the torsion angles of the 2-coordinate nitrogen atoms. For instance, the group of structures with enthalpies from -25.0 to -25.5~eV/FU exhibit a more crumpled N layer in comparison to those found between -28.3 to -29.0~eV/FU, which possess a relatively flat 2D N layer. The unconstrained EA search results in a larger diversity of optimized structures, which is reflected by their greater range in enthalpies, as depicted in Figure \ref{fig:xen8}(a). Specifically, 96.0\% of the low-lying structures are located within the enthalpy range of -18.0 to -30.0~eV/FU. The blue or green jagged lines in the two searches highlight the enthalpy of the most stable phase found, and the red stars the first instance that the target phases were identified. 

The ground state XeN$_{8}$ phase, whose unit cell is illustrated in Figure \ref{fig:xen8}(c), features exclusively N$_2$ dumbbells that surround the Xe atoms, and it possesses $P$-1 symmetry. It was the 713$^{th}$ structure to be generated during the course of the unconstrained EA search (Figure \ref{fig:xen8}(a)), and was found four times within the run. Because N$_2$ molecules comprise this polymorph, it cannot be considered as a candidate for a HEDM. An alternative XeN$_8$ crystal that contains 2D polymeric nitrogen motifs could be based upon intercalating the noble gas atom into the pores of 18-crown-6 N$_{18}$ units, whose Lewis structure is presented in Figure~\ref{fig:xen8}(b, inset)\cite{wang2022putting}. It is noteworthy that the 2D polynitrogen layer based upon the 18-crown-6 N$_{18}$ motifs exhibits a specific ratio of nitrogen atoms with different coordination numbers:  1/3 of the nitrogen atoms are three-coordinate (highlighted by yellow dots in Figure \ref{fig:xen8}(b), while 2/3 are two-coordinate. Each ring of the aromatic 2D polynitrogen layer contains ten $\pi$ electrons per N$_{18}$ unit. These electrons are distributed such that each N$^{t}$ (tri-coordinate) atom has one $\pi$ lone pair shared by three rings (6 $\times$ 1/3 $\pi$ pairs), and each N$^{d}$ (di-coordinate) atom has one $\pi$ N$^{d}$=N$^{d}$ pair shared by two rings (6 $\times$ 1/2 $\pi$ pairs). This molecular orbital configuration follows the H$\ddot{u}$ckel rule and results in a 4$n$ + 2 electron count, which confers chemical stability to the 2D polynitrogen layer. 

What type of constraints could be used to find a metastable crystal lattice based upon this N$_{18}$ motif? As alluded to above, one possibility could be to constrain the coordination numbers of the nitrogen atoms. Requiring 1/3 of the N atoms to be 3-coordinate and the remainder to be 2-coordinate in the breeding pool of a CSP search performed at 50~GPa resulted in the discovery of the $P$-1 symmetry phase illustrated in Figure \ref{fig:xen8}(d), which is consistent with previous predictions~\cite{wang2022putting}. It is noted that the predicted Phase $P$-1 II has not yet been synthesized. As desired, this phase contains slightly puckered 2D polynitrogen layers in the $bc$ plane that are separated by Xe atoms along the $a$ lattice vector, and the layers are stacked directly on top of each other.  At 50~GPa this desired $P$-1 II phase is computed to be 0.918~eV/FU higher in enthalpy than the $P$-1 I phase, but by 100~GPa it becomes the ground state XeN$_{8}$ polymorph. Despite the fact that $P$-1 II  XeN$_8$ remains dynamically stable, and could be kinetically trapped upon decompression to ambient pressure, it becomes increasingly thermodynamically disfavored at lower pressures, where structures featuring N$_2$ molecular units, like $P$-1 I, dominate. Indeed, $P$-1 II was not predicted in the traditional search illustrating that without imposing the aforementioned constraints, this metastable phase would never be found at 50 GPa. Moreover, no common structure is predicted in the traditional and constrained searches. This example highlights the effectiveness of our constrained search method in isolating a specific metastable phase, here, based on coordination number, among a large number of possible configurations. Though this EA search was performed using the PBE functional, enthalpy comparisons based upon functionals that include Van der Waals (vdW) dispersion yielded the same trends (see Table S3 in SI).

\subsection{Finding Polymorphs through Bravais Lattice Constraints}
\begin{figure}[]
\centerline{\includegraphics[width=0.6\columnwidth]{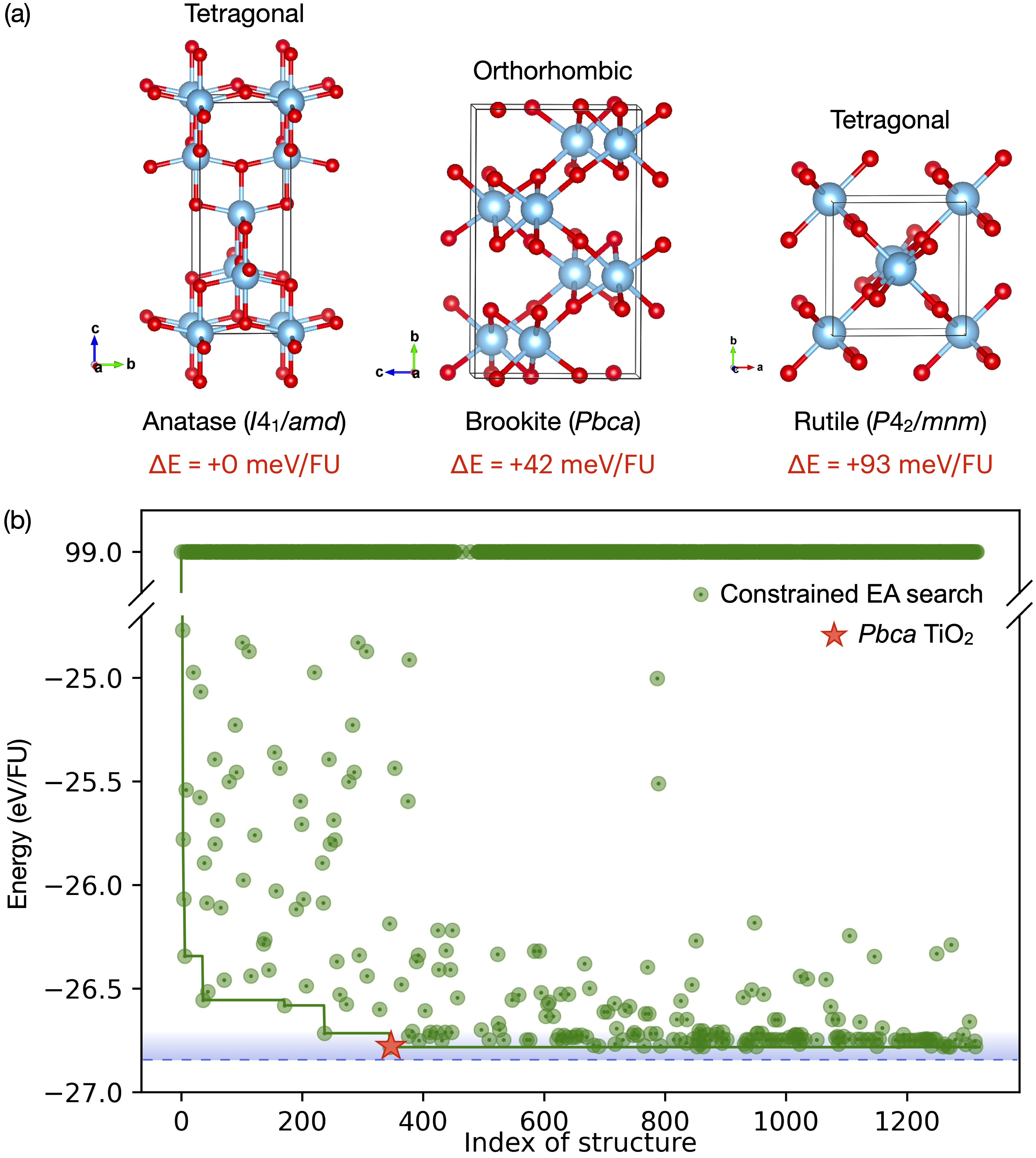}}
\caption{(a) The crystal structures of the three main TiO$_2$ phases: anatase ($I$4$_{1}$/$amd$ space group), brookite ($Pbca$ space group), rutile ($P$4$_{2}$/$mnm$ space group). Their PBE-calculated relative enthalpies are provided, with anatase being the most stable. (b) The evolution of the enthalpy versus the structure index for TiO$_{2}$ at 0 GPa in a search constrained so that the breeding pool only contains structures with an orthorhombic Bravais lattice. The 345$^{th}$ structure corresponded to the most stable phase satisfying this constraint, as indicated by the red star. The blue dashed line represents the calculated energy of the ground state (anatase, $I$4$_{1}$/$amd$ TiO$_{2}$). Oxygen atoms are red and titanium atoms are blue balls.}
\label{fig:tio2}
\end{figure}


In addition to considering the local crystalline order (represented by coordination number for the test case of XeN$_{8}$), unit cell symmetry could also be leveraged as a criterion for predicting the crystal structure of a desired metastable compound. Here, we take TiO$_{2}$ as an example, aiming to predict the polymorphs that possess an orthorhombic Bravais lattice. Multiple phases of TiO$_{2}$ are known with the three main polymorphs being anatase (FU = 4), rutile (FU = 2), and brookite (FU = 8)~\cite{trail2017quantum,guo2019fundamentals}, whose crystal structures are shown in Figure~\ref{fig:tio2}(a). Using the PBE functional at 0~GPa anatase turns out to be the ground state~\cite{trail2017quantum}, though various interatomic potentials have predicted that either rutile or brookite are preferred \cite{woodley2009structure}. Within the PBE functional, which we use here, brookite and rutile turned out to be metastable with their energies being 42 and 93~meV/FU higher than that of anatase, respectively. As such, an unconstrained \textsc{XtalOpt} search would be expected to yield anatase TiO$_2$ as the lowest energy phase. 

In all three of these polymorphs the titanium atoms are six coordinated either in a trigonal prismatic (brookite), or a perfect octahedral (rutile) or distorted octahedral (anatase) configuration. The oxygen atoms, on the other hand, exhibit a 3-coordinate arrangement in all forms of rutile, anatase, and brookite. Therefore, if our goal was to uncover the brookite phase using a CSP search, constraining the coordination numbers on the titanium and oxygen atoms might not yield the desired target. Could the crystal lattices of these polymorphs be used instead? While both anatase (space group $I$4$_{1}$/$amd$) and rutile (space group $P$4$_{2}$/$mnm$) adopt a tetragonal lattice, brookite (space group $Pbca$) possesses an orthorhombic Bravais lattice.  One of \textsc{Pymatgen}'s 
libraries allows us to obtain the Bravais lattice for a crystal structure, and can be used, as described in Section S1 in the SI, to constrain the breeding pool. Consequently, in an attempt to specifically locate the orthorhombic brookite phase, we carried out a constrained evolutionary search in which exclusively those structures with an orthorhombic Bravais lattice were retained in the pool, while all other structures were eliminated. 

Figure \ref{fig:tio2}(b) illustrates the evolution of the calculated enthalpy as a function of the structure index for TiO$_{2}$ at 0~GPa in a constrained search. In this search, wherein around 1300 structures were generated and optimized, brookite appeared as the 345$^{th}$ crystal and was highlighted by the red star. 
An analogous unconstrained search was conducted, and during the course of the \textsc{XtalOpt} run a running list of all of the structures analysed, ranked from lowest to highest in enthalpy or energy, was maintained. The contents of this output file, named `results.txt', for the unconstrained search is presented in Section S4 in the SI. The unconstrained search ranked anatase, the 187$^{th}$ crystal found, as the most stable polymorph with the highest rank, whereas brookite and rutile were found as the 1141$^{th}$ and 559$^{th}$ structures with rank \#16 and \#25, respectively, see Figure S2 in the SI. Though both the constrained and unconstrained searches identified brookite, the former was much more efficient finding it as the 345$^{th}$ as compared to the 1141$^{th}$ structure. This indicates the inadequacy of the conventional evolutionary search process when compared to the constrained approach for identifying a specific metastable phase. 

\subsection{Sublattice Symmetry for Aiding Experimental Structural Determination}
In contrast to previous studies \cite{wales1997global,wheeler2007sass,huber2023targeting},  which specifically target high-symmetry metastable structures, our approach can enforce structural criteria that is not restricted to high-symmetry groups but can incorporate any information provided by experimental insights.Though XRD is an invaluable tool for the experimental characterization of a crystalline material, sometimes it cannot be used alone to determine a compound's structure. This is particularly true when elements with similar scattering factors comprise the phase (\eg, it is difficult to distinguish between boron and carbon via XRD), heavy elements with very strong scattering factors mask signals arising from weaker scatters, or when light elements -- hydrogen being the most notorious -- are present. Another example is high pressure research where XRD patterns obtained in either static~\cite{pena2022chemically,dasenbrock2023evidence} or dynamic compression~\cite{polsin2022structural,wang2023topological} experiments may only provide incomplete data due to the experimental set-up.

\begin{figure}[]
\centerline{\includegraphics[width=0.6\columnwidth]{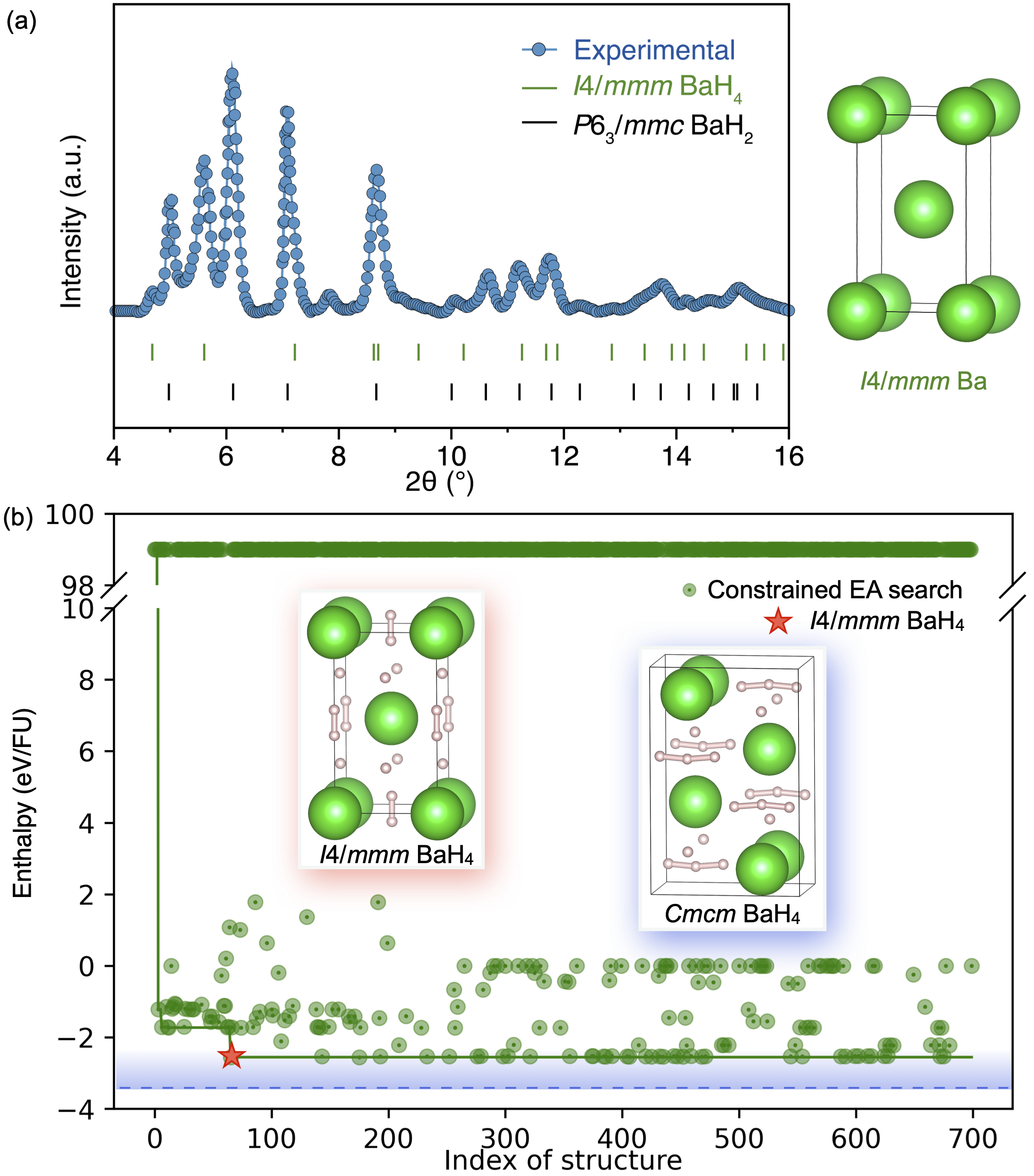}}
\caption{(a) The XRD pattern of a phase mixture of barium polyhydrides at 45~GPa from Ref\cite{pena2022chemically}. The tick marks indicate Bragg peaks from both $I$4/$mmm$ BaH$_{4}$ and $P$6$_{3}$/$mmc$ BaH$_{2}$. The right panel displays the unit cell of an $I$4/$mmm$ Ba configuration. (b) The evolution of the enthalpy versus the structure index for BaH$_{4}$ at 50~GPa in an \textsc{XtalOpt} search constrained so that the breeding pool only contained structures with an $I$4/$mmm$ symmetry Ba sublattice. The 65$^{th}$ structure corresponded to the most stable phase satisfying this constraint, as indicated by the red star.  The blue dashed line represents the calculated enthalpy of the ground state ($Cmcm$ BaH$_{4}$), which was found in an unconstrained search. The crystal structures of the $I$4/$mmm$ and $Cmcm$ BaH$_{4}$ phases are inserted in (b). Barium atoms are green and hydrogen atoms white balls.}
\label{fig:bah4}
\end{figure}

In such situations, an EA search constrained by the available XRD data may be extremely useful in uncovering the structure of a synthesized phase. In fact, in the field of high pressure research unconstrained CSP searches have already proven to be invaluable in predicting targets for synthesis and identifying the materials that have been made~\cite{zurek2015predicting}. A case in point is a recent report that described the synthesis of BaH$_{4}$ using high-pressure diamond anvil cell experiments\cite{pena2022chemically}. Synchrotron measurements were performed to obtain the XRD pattern of the synthesized polyhydride, with the experimental XRD pattern depicted in Figure \ref{fig:bah4}(a). Analysis of the data led the researchers to conclude that it resulted from a co-existence of two phases. One of these was a $P$6$_{3}$/$mmc$ BaH$_{2}$ compound, which was previously known \cite{pena2021synthesis}, and the second was attributed to an unknown BaH$_4$ phase. Given the weak scattering power of hydrogen atoms~\cite{schmidtmann2014determining,cheng2017synchrotron}, the symmetry that was associated with the second unknown phase, $I$4/$mmm$, could only provide information about the structure of the heavy barium sublattice (see inset). Though the hydrogen atom positions could not be extracted directly from the experimental data, their assignment could be made based upon DFT calculations that relied on CSP combined with analogy to other known structures. The ground state found via CSP performed using the \textit{ab initio} random structure searching (AIRSS)\cite{pickard2009structures} method was a $Cmcm$ symmetry crystal~\cite{pena2022chemically} that did not quite reach the Ba-H convex hull. However, the experiment was in better agreement with a metastable phase whose enthalpy lay $\sim$300~meV/FU above the ground state, which was isotypic with many of the tetrahydrides of an electropositive metal element that have been predicted or synthesized under pressure~\cite{Zurek:2020a}. These two structures are compared in the inset in Figure \ref{fig:bah4}(b). To determine if a more stable phase could explain the experimental data, it is imperative to perform a search that constrains the breeding pool to $I$4/$mmm$ Ba sublattices.

To illustrate the utility of a constrained \textsc{XtalOpt} search in such a situation, 
one of \textsc{Pymatgen}'s libraries was employed to determine the space group of the Ba sublattice of each optimized structure, see Section S1 in the SI. Figure \ref{fig:bah4}(b) illustrates the evolution of the calculated enthalpy as a function of the structure index. The phase with the lowest enthalpy identified in this search, with an $I$4/$mmm$ space group, first appeared as the 65$^{th}$ structure, indicated with a red star in Figure \ref{fig:bah4}(b). It was identical to the $I$4/$mmm$ BaH$_4$ phase proposed in Ref.\cite{pena2022chemically}, and isotypic with many of the previously predicted or synthesized metal tetrahydrides~\cite{Zurek:2020a}, confirming the structural assignment made in Ref.\ \cite{pena2022chemically}.

Our DFT calculations found that the enthalpy of $I$4/$mmm$ BaH$_{4}$ was 277~meV/FU higher than that of the $Cmcm$ phase, which appeared as the 375$^{th}$ structure in an unconstrained search, confirming it was the ground state. In the unconstrained EA run $I$4/$mmm$ BaH$_{4}$ was ranked as the 354$^{th}$ lowest enthalpy candidate in the `results.txt' file, and it is therefore unlikely that it would have been chosen for further analysis from an unconstrained search without the availability of experimental data. This highlights the inefficiency of the traditional evolutionary search method, as compared to the constrained one, for identifying a specific metastable phase. The large enthalpy difference between the $I$4/$mmm$ and $Cmcm$ tetrahydrides substantiates the hypothesis that the former could be  an intermediate phase resulting from hydrogen migration through the Ba lattice\cite{pena2022chemically}. 

Furthermore, it is worth mentioning that the methodology described in this section was used by us in an attempt to uncover the structure of an unknown-stoichiometry N-doped lutetium hydride \cite{hilleke2023structure}, which was suggested to comprise the putative room-temperature superconductor recently reported under kbar pressures \cite{dasenbrock2023evidence}. The measured XRD pattern could only provide evidence about the Lu sublattice, which was proposed to be face-centered-cubic ($fcc$) and did not reveal the positions of the hydrogen or nitrogen atoms. In our work several metastable Lu-N-H structures featuring $fcc$ Lu were successfully identified using constrained \textsc{XtalOpt} searches, although their estimated $T_c$s were far below room temperature~\cite{hilleke2023structure}.

\subsection{Molecular Units by Constraining the Chemical Environment}
\begin{figure}[]
\centerline{\includegraphics[width=0.7\columnwidth]{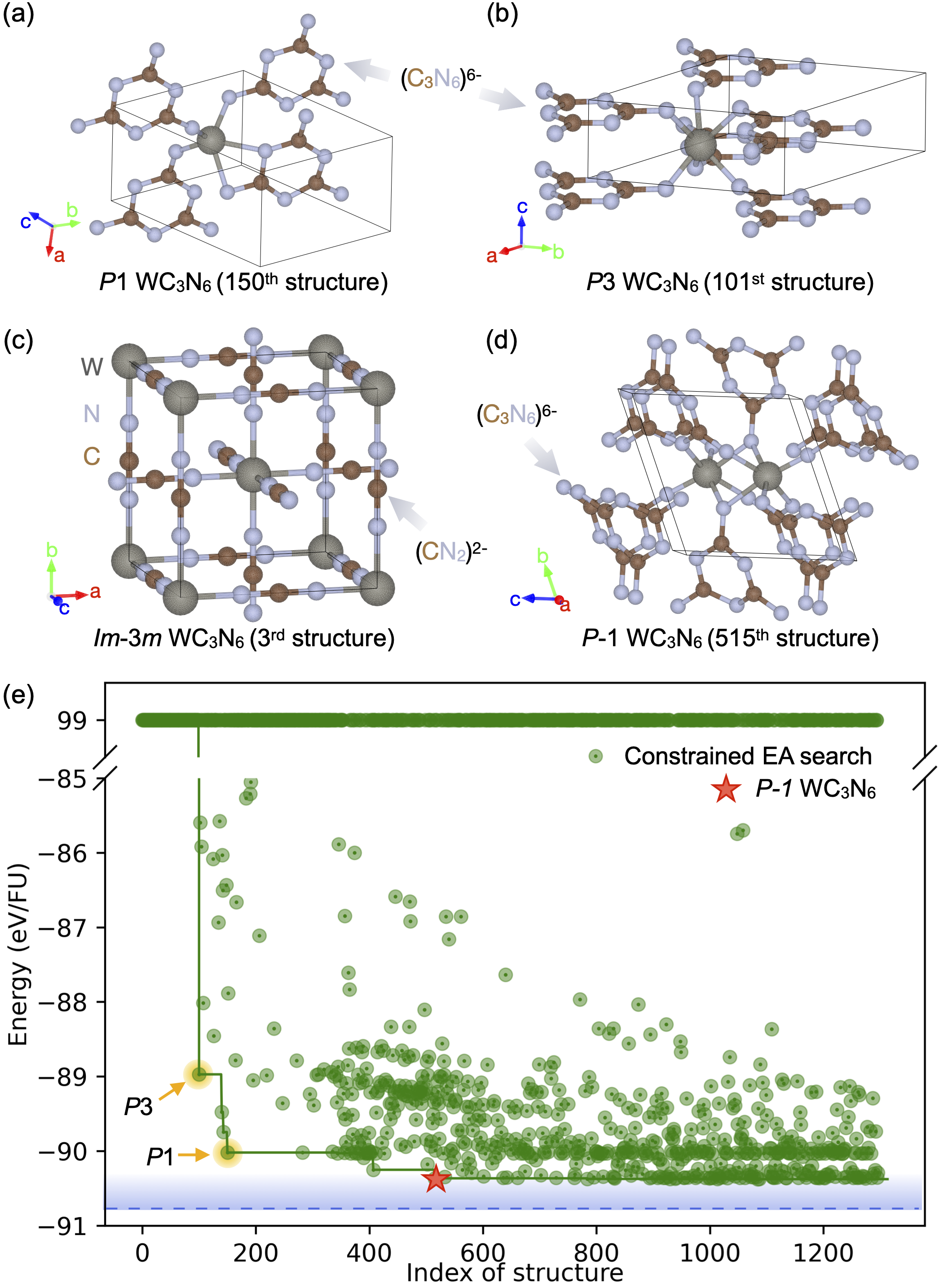}}
\caption{(a-d) The crystal structures of multiple phases of WC$_{3}$N$_{6}$. (a, b) The previously proposed $P$1 and $P$3 polymorphs with (C$_{3}$N$_{6}$)$^{6-}$ units~\cite{chen2023computational}. (c) The $Im$-3$m$ phase, which is the ground state, found in a traditional search. This compound contains W$^{6+}$ octahedrally coordinated by [N=C=N]$^{2-}$ units. (d) The most stable $P$-1 WC$_{3}$N$_{6}$ salt-like phase containing (C$_{3}$N$_{6}$)$^{6-}$ motifs found in a constrained search. (e) The evolution of the energy versus structure index in an ambient pressure constrained CSP search for WC$_{3}$N$_{6}$. The 515$^{th}$ structure with space group $P$-1 is indicated by the red star, while the 150$^{th}$ and 101$^{st}$ structures (seeds) are indicated by the orange arrows. The blue dashed line represents the calculated energy of the ground state ($Im$-3$m$ WC$_{3}$C$_{6}$) from an unconstrained search. The energies of the discarded phases in the constrained search are represented as a constant value of 99~eV/FU.}
\label{fig:wc3n6}
\end{figure}


In a search for new photocatalysts for water splitting, Chen and coworkers used CSP methods to uncover two energetically favorable polymorphs of WC$_{3}$N$_{6}$, a salt comprised of (C$_{3}$N$_{6}$)$^{6-}$, a fully deprotonated form of melamine (C$_{3}$H$_{6}$N$_{6}$) \cite{chen2023computational}. The two polymorphs, which adopt space groups $P$1 and $P$3, were predicted using the particle swarm optimization algorithm \cite{wang2010crystal}. Within them, the tungsten atom is found in either a 5-coordinated ($P$1) or 6-coordinated ($P$3) environment, as depicted in Figure \ref{fig:wc3n6}(a,b). Given that the  melaminate anion, (C$_{3}$N$_{6}$)$^{6-}$, has a higher energy than three carbodiimide anions, (CN$_{2}$)$^{2-}$, we wondered if a constrained \textsc{XtalOpt} search would be successful in locating novel polymorphs of  WC$_{3}$N$_{6}$?

The constraints that were the most appropriate for our target could be derived from the chemical environments and coordination numbers of C and N within the melaminate anion. Specifically, each C atom is coordinated to three N atoms, and whereas half of the N atoms are expected to be coordinated to a single C atom, the remaining half will be coordinated to two C atoms, see the melaminate in Figure \ref{fig:wc3n6}(a-b). The cutoff value for the C-N bond distance of 1.5~\AA{} was chosen based upon a comparison with the expected single and double bond lengths between carbon and nitrogen, as discussed more fully in the SI. The current version of \textsc{XtalOpt} is unable to create the first generation of structures by making use of molecular units, unless they conform to the well-known valence shell electron pair repulsion (VSEPR) geometries. Since the melaminate anion cannot be represented by a VSEPR geometry, it is highly unlikely that the structures generated via this method or using the \textsc{RandSpg} algorithm would possess suitable chemical environments. Therefore, in addition to utilizing the aforementioned constraints to filter the breeding pool, the initial generation was populated with 100 random structures generated by \textsc{RandSpg}, the previously predicted $P$1 and $P$3 configurations \cite{chen2023computational}, and with 98 molecular crystals of WC$_{3}$N$_{6}$ generated by the \textsc{PyXtal} code \cite{fredericks2021pyxtal}. The initial generation was created in this way for both the constrained and traditional searches. It should be emphasized that structural seeds were not used in any of the other three systems that were studied in this work.

The most stable phase found in the unconstrained search was a carbodiimide-like salt in the $Im$-3$m$ ($Z=2$) space group featuring  W(VI) and 3 (CN$_{2}$)$^{2-}$ moieties (Figure \ref{fig:wc3n6}(c)), which was found in the initial random generation as the third structure. Though we did not locate any other polymorphs in the literature with the WC$_{3}$N$_{6}$ stoichiometry featuring carbodiimide anions, a W$_2$(CN$_{2}$)$_3$ phase with these units, which is isostructural with an experimentally known $R$-3$c$ Cr$_2$(CN$_{2}$)$_3$\cite{tang2010ferromagnetic} compound has been reported in the Materials Project database~\cite{jain2013commentary}. 

In the constrained search, however, the most stable structure located corresponded to a $P$-1 ($Z=2$) WC$_{3}$N$_{6}$ metastable phase with two W atoms in a distorted octahedral coordination, as shown in Figure \ref{fig:wc3n6}(d), with an average W-N distance of 2.04~\AA{}. This distance is extremely close to the expected W-N bond length of 2.06~\AA{} for a W$^{6+}$ ion coordinated to N$^{3-}$ in a sixfold coordination state \cite{chen2023computational}. Phonon calculations confirmed that this structure was a local minimum. The previously reported $P$3 and $P$1 structures were found as the 101$^{st}$ and 150$^{th}$ phases in the constrained search (since they were employed as seeds), as illustrated in the energy versus structure index plot in Figure \ref{fig:wc3n6}(d), and the newly predicted polymorph was found as the 515$^{th}$ structure. The energy of the new $P$-1 phase was calculated to be 53~meV/atom higher than the global minimum, $Im$-3$m$ W(CN$_{2}$)$_{3}$, but 29~meV/atom (28 kJ mol$^{-1}$) lower than the previously reported metastable $P$1 phase at the D3 (BJ) level of theory.  The relative stability of these phases has also been confirmed by the optB88 and PBE theories as displayed in Table \ref{tbl:wc3n6-energy}. It is worth mentioning that the energy difference between $P$3 and $P$1 WC$_{3}$N$_{6}$ is calculated to be 100 meV/atom (97 kJ mol$^{-1}$), consistent with the value of 96 kJ mol$^{-1}$ reported in Ref. \cite{chen2023computational}. As discussed in Section S5 of the SI, a search that did not employ the previously predicted $P$3 and $P$1 structures as seeds was also able to find the novel $P$-1 phase, with similar efficiency to the search whose results are shown in Figure \ref{fig:wc3n6}(e).

\begin{table}
	\centering
	\caption{The relative energies of various WC$_{3}$N$_{6}$ phases as calculated at the D3 (BJ), optB88, and PBE levels of theory.}
	\label{tbl:wc3n6-energy}
	\begin{threeparttable}
	\setlength{\tabcolsep}{7mm}{
	\begin{tabular}{cccc}
		\hline
		\hline
		\multirow{2}{*}{Phases}  & \multicolumn{3}{c} {$\Delta\emph{E}$ (meV/atom)}  \\
				                 & D3 (BJ) & optB88  & PBE    \\   
		\hline
		$Im$-3$m$                &   -53  & -42  & -45 \\
		$P$-1\tnote{a}           &   0    &  0   &  0  \\
		$P$1\tnote{b}            &  29    &  56  &  11 \\
		$P$3\tnote{b}            &  129   &  145 &  120 \\
		\hline
		\hline
	\end{tabular}
	\begin{tablenotes}
		\footnotesize
		\item[a]This work. 
		\item[b]Ref \cite{chen2023computational} .
	\end{tablenotes}}
	\end{threeparttable}
\end{table}

\section{Conclusions}
A method to predict metastable phases with specific structural features has been developed and implemented through a constrained scheme that utilizes the \textsc{XtalOpt} software and the \textsc{Pymatgen} library. This approach involves automating global crystal prediction algorithms so the structures kept in their breeding pools meet user defined criteria pertaining to either local crystalline order or symmetry. The accuracy and effectiveness of this method was evaluated by benchmarking its performance against three metastable systems: XeN$_{8}$ with 2D polymeric nitrogen motifs at 50~GPa, brookite TiO$_{2}$ at ambient pressure, and a synthesized BaH$_{4}$ phase that exhibits an $I$4/$mmm$ Ba sub-lattice, as deduced via experimental X-ray diffraction (XRD) measurements at 50~GPa. Additionally, we have predicted a new metastable melaminate salt, $P$-1 WC$_{3}$N$_{6}$, whose energy was found to be lower than a recently reported phase \cite{chen2023computational}. 

Our study demonstrates that a constrained search is more efficient in identifying metastable phases with predefined structural features as compared to a traditional crystal structure search. The constrained method presented herein offers a substantial degree of flexibility, as it permits users to develop constraint criteria according to their specific requirements, and with the help of shared scripts provided in the SI. In situations where experimental data is available, infrared or Raman data may suggest coordination number constraints, while XRD data might indicate constraints related to symmetry, like the Bravais lattice (TiO$_{2}$ case) or space group of the crystal, or even of its sub-lattice (BaH$_{4}$ case). This is especially relevant in high-pressure conditions where it may not be possible to fully measure the structural characteristics of a synthesized compound, and only partial geometric details can be determined in such circumstances. Additionally, based on chemical bonding rules, the local crystalline order (XeN$_{8}$ case) and chemical environment (WC$_{3}$N$_{6}$ case), which are associated with specific material properties, could serve as the constraint criteria for the discovery of advanced materials with desired functionalities. The methodology described here could be used in conjunction with other CSP codes, illustrating its broad impact. We believe this method has the potential to aid in the identification of experimental compounds and the design of metastable phases important for basic science and for a broad range of potential applications.

\newpage
\section{Associated Content}
The Supporting Information can be accessed at no cost through https://pubs.acs.org/doi/xxx.\\ This resource provides comprehensive information on the constrained search methodology employed in conjunction with \textsc{XtalOpt} and \textsc{Pymatgen}. It includes details on the associated scripts, the workflow of a conventional EA search, the criteria for search convergence, and the computational complexity. Furthermore, the calculated structural parameters for the phases discussed are presented therein. 
Additionally, the dynamic stability of the $P$-1 WC$_{3}$N$_{6}$ phase was investigated.

\section{Author Information}
\subsubsection{Corresponding Author}
Eva Zurek - Department of Chemistry, State University of New York at Buffalo, Buffalo, NY 14260-3000, USA; ORCID: 0000-0003-0738-867X\\
Email: ezurek@buffalo.edu

\subsubsection{Authors}
Busheng Wang - Department of Chemistry, State University of New York at Buffalo, Buffalo, NY 14260-3000, USA; ORCID: 00000002-7743-9471\\
Katerina P. Hilleke - Department of Chemistry, State University of New York at Buffalo, Buffalo, NY 14260-3000, USA; ORCID: 0000-0003-4322-8403\\
Samad Hajinazar - Department of Chemistry, State University of New York at Buffalo, Buffalo, NY 14260-3000, USA; ORCID: 0000-0002-7255-5932\\
Gilles Frapper - Applied Quantum Chemistry Group, E4 Team, IC2MP UMR 7285, Université de Poitiers-CNRS, 86073 Poitiers, France; ORCID: 0000-0001-5177-6691

\subsubsection{Author Contributions}
E.Z., G.F.\ and B.W., conceived the research. B.W.\ carried out the calculations and E.Z.\ supervised the study. K.H.\ and S.H.\ contributed to the analysis and made improvements to the \textsc{XtalOpt} code. All authors participated in discussing the results. The manuscript was written in collaboration with all authors. All authors have given approval to the final version of the manuscript.

\subsubsection{Notes}
The authors declare no competing financial interest.

\section{Acknowledgements:}

We acknowledge the U.S. National Science Foundation (DMR-2136038 and DMR-2119065) for financial support. K.H.\ acknowledges the Chicago/DOE Alliance Center under Cooperative Agreement Grant No.\ DE-NA0003975. Calculations were performed at the Center for Computational Research at SUNY Buffalo (USA)\cite{ccr}. Gilles Frapper acknowledges the High-Performance Computing Center of Joliot-Curie Rome/TGCC GENCI (France) under Project No. A0140807539. \\

\newpage
\bibliography{bibliography.bib}

\newpage
For Table of Contents Only
\begin{figure}[]
\centerline{\includegraphics[width=8.5cm]{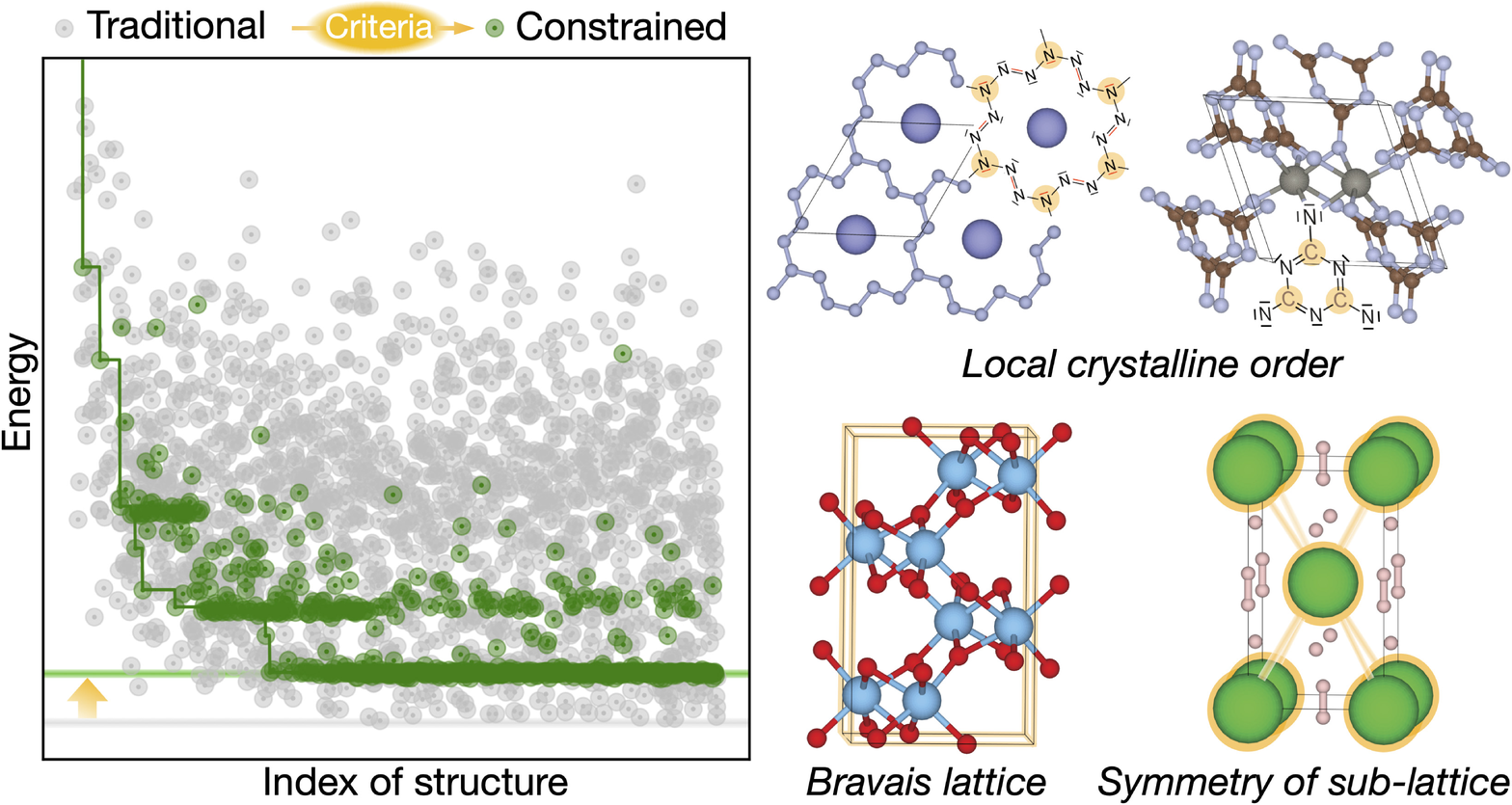}}
\label{fig:toc}
\end{figure}

\end{document}